\documentclass{aa}  
\newcommand{\HA}{H$\alpha$}
\newcommand{\HB}{H$\beta$}
\newcommand{\HG}{H$\gamma$}
\newcommand{\kms}{km\,s$^{-1}$}
\newcommand{\degree}{^{\circ}}%
\usepackage{graphicx}
\usepackage[normalem]{ulem}
\usepackage{textcomp}
\usepackage{hyperref}

\makeatletter
\renewcommand*\aa@pageof{, page \thepage{} of \pageref*{LastPage}}
\makeatother
\usepackage{txfonts}
\usepackage{lipsum}
\hypersetup{draft}

\begin{document}

   \title{The quest for stellar coronal mass ejections in late-type stars}

   \subtitle{I. Investigating Balmer-line asymmetries of single stars in Virtual Observatory data}

   \author{Kriszti\'an Vida
          \inst{1}
          \and
          Martin Leitzinger\inst{2,3}
          \and
          Levente Kriskovics\inst{1}
          \and
          B\'alint Seli\inst{1,4}
          \and
          Petra Odert\inst{2,3}
          \and
          Orsolya Eszter Kov\'acs\inst{1,4,5}
          \and
          Heidi Korhonen\inst{6}
          \and
          Lidia van Driel-Gesztelyi\inst{1,7,8}
          }

   \institute{
   Konkoly Observatory, MTA CSFK, H-1121 Budapest, Konkoly Thege M. \'ut 15-17, Hungary\\
   \email{vidakris@konkoly.hu}
   \and
   Institute of Physics/IGAM, University of Graz, Universit\"atsplatz 5, A-8010 Graz, Austria
   \and
   Space Research Institute, Austrian Academy of Sciences, Schmiedlstra\ss{}e 6, A-8042 Graz, Austria
   \and
   E\"otv\"os University, Department of Astronomy, Pf. 32, 1518 Budapest, Hungary 
   \and
   Harvard Smithsonian Center for Astrophysics, 60 Garden Street, Cambridge, MA 02138, USA
   \and
   Dark Cosmology Centre, Niels Bohr Institute, University of Copenhagen, Juliane Maries Vej 30, DK-2100 Copenhagen, Denmark
   \and
   Mullard Space Science Laboratory, University College London, Holmbury St. Mary, Dorking, UK
   \and
   LESIA, Observatoire de Paris, Universit\'e PSL, CNRS, Sorbonne Universit\'e, Univ. Paris Diderot, Sorbonne Paris Cit\'e, Meudon, France
   }

   \date{Received Sept 18, 2018; accepted \dots, 2018}

 
  \abstract
   {Flares and coronal mass ejections (CMEs) can have deleterious effects on their surroundings: they can erode or completely destroy atmospheres of orbiting planets over time and also have high importance in stellar evolution. Most of the CME detections in the literature are single events found serendipitously sparse for statistical investigation.}
   {We aimed to gather a large amount of spectral data of M-dwarfs  to drastically increase the number of known events to  make statistical analysis possible in order to study the properties of potential stellar CMEs.}
   {Using archive spectral data  we investigated asymmetric features of Balmer-lines, that could indicate the Doppler-signature of ejected material.}
   {Of more than 
   5500 spectra we found 478 with line asymmetries -- including nine larger events, in terms of velocity and mass--on 25 objects,  with 1.2--19.6 events/day on objects with line asymmetries.
   Most events are connected with enhanced peak of Balmer-lines, indicating that these are connected to flares similar to solar events. In most cases the detected speed does not reach surface escape velocity: the typical observed maximum velocities are in the order of 100--300\,\kms , while the typical masses of the ejecta were in the order of $10^{15}-10^{18}$g.
   Statistical analysis of the events suggests that these events are more frequent on cooler stars with stronger chromospheric activity.
   }
   {If the detected events correspond to CMEs, the detected maximum velocities are lower than those observed on the Sun, while event rates were somewhat lower than we could expect from the solar case. If the velocities are not distorted significantly due to a projection effect, these findings may support the idea that most of the coronal mass ejections could be suppressed by strong magnetic field. Alternatively, it is possible that we can observe only an early low-coronal phase of the events before being accelerated at higher altitudes. {Our findings could indicate that later-type,  active  dwarfs could  be  a  safer  environment  for  exoplanetary  systems CME-wise than previously thought, and atmosphere loss due to radiation effects would play a stronger role in exoplanetary atmosphere evolution than CMEs.}}


   \maketitle
%

\section{Introduction} 
\begin{table*}
\caption{Summary of the archival spectra used for our analysis.}
\label{tab:vospec}
\centering
\begin{tabular}{lc|lc|lc|lc|lc}
\hline\hline
star	&	number of	&	star	&	number of	&	star	&	number of	&	star	&	number of	&	star	&	number of\\
&	spectra	&&	spectra	&&	spectra	&&	spectra	&&	spectra\\
\hline
GJ\,47	&	  8	&	GJ\,273	&	 24	&	GJ\,503.2&	 21	&	GJ\,890		&	350	&	GJ\,3801	&	  8\\
GJ\,48	&	  8	&	GJ\,285	&	177	&	GJ\,514	&	  6	&	GJ\,905		&	 19	&	GJ\,3967	&	 24\\
GJ\,49	&	181	&	GJ\,299	&	  4	&	GJ\,526	&	  9	&	GJ\,908		&	 19	&	GJ\,3971	&	 24\\
GJ\,51	&	122	&	GJ\,317	&	  8	&	GJ\,536	&	  9	&	GJ\,1105	&	  8	&	GJ\,4040	&	  8\\
GJ\,70	&	 16	&	GJ\,357	&	  8	&	GJ\,555	&	 12	&	GJ\,1111	&	117	&	GJ\,4053	&	 16\\
GJ\,83.1&	 30	&	GJ\,382	&	  7	&	GJ\,559.1&	225	&	GJ\,1125	&	  8	&	GJ\,4070	&	  8\\
GJ\,96	&	  9	&	GJ\,386	&	 12	&	GJ\,581	&	 10	&	GJ\,1148	&	  9	&	GJ\,4071	&	 24\\
GJ\,109	&	  8	&	GJ\,388	&	217	&	GJ\,625	&	 15	&	GJ\,1154	&	  8	&	GJ\,4247	&	120\\
GJ\,117	&	  4	&	GJ\,393	&	 19	&	GJ\,628	&	  2	&	GJ\,1156	&	 93	&	GJ\,4333	&	  8\\
GJ\,123	&	  2	&	GJ\,394	&	  5	&	GJ\,673	&	  1	&	GJ\,1167	&	  4	&	GJ\,9520	&	287\\
GJ\,154	&	  2	&	GJ\,406	&	 25	&	GJ\,686	&	  7	&	GJ\,1224	&	 61	&	HD\,77407	&	  4\\
GJ\,170	&	 24	&	GJ\,408	&	 21	&	GJ\,694	&	 11	&	GJ\,1243	&	 26	&	HD\,189733	&	268\\
GJ\,172	&	  2	&	GJ\,410	&	340	&	GJ\,729	&	 98	&	GJ\,1245	&	155	&	HD\,209458	&	 65\\
GJ\,173	&	  4	&	GJ\,411	&	 54	&	GJ\,735	&	108	&	GJ\,1289	&	 10	&	HIP\,103039	&	  8\\
GJ\,176	&	  2	&	GJ\,424	&	125	&	GJ\,793	&	  8	&	GJ\,2066	&	  8	&	LHS\,2613	&	  1\\
GJ\,179	&	  9	&	GJ\,431	&	 44	&	GJ\,803	&	 65	&	GJ\,3126	&	  8	&	LHS\,2686	&	  9\\
GJ\,192	&	  9	&	GJ\,436	&	 15	&	GJ\,816	&	  8	&	GJ\,3323	&	  2	&	LTT\,763	&	  1\\
GJ\,205	&	 55	&	GJ\,445	&	  4	&	GJ\,821	&	  4	&	GJ\,3378	&	  8	&	V363\,Lac	&	  4\\
GJ\,208	&	  1	&	GJ\,447	&	  9	&	GJ\,825	&	  8	&	GJ\,3459	&	  4	&	$\epsilon$\,Eri	&	235\\
GJ\,212	&	  2	&	GJ\,480	&	  8	&	GJ\,873	&	212	&	GJ\,3622	&	 80	&	$\kappa^1$\,Cet	&	 13\\
GJ\,213	&	  8	&	GJ\,486	&	  8	&	GJ\,875.1&	 17	&	GJ\,3647	&	 54	&	$\chi^1$\,Ori	&	481\\
GJ\,226	&	  8	&	GJ\,493.1&	 20	&	GJ\,876	&	 10	&	GJ\,3780	&	  2	&   \\
GJ\,251	&	 47	&	GJ\,494	&	281	&	GJ\,887	&	  4	&	GJ\,3789	&	 20	&	\\

\hline
\end{tabular}
\end{table*}

Flares and coronal mass ejections (CMEs) are some of the most prominent, most energetic events for stellar activity. 
These events can have high importance for stellar evolution, but also in exoplanet studies: frequent high energy events could strip off or erode the atmospheres of nearby orbiting planets, rendering them uninhabitable \citep{2007AsBio...7..167K,2008SSRv..139..437Y}. CMEs can also alter the atmospheres of the exoplanets -- if these have high enough energy and are frequent, the planetary atmospheres will be continuously altered, which is disadvantageous for hosting life (see \citealt{trappist1} and references therein).
On the Sun, CMEs are studied in high detail, both by observation and modeling (see the review of \citealt{cme-livingreview, livingreview}, and references therein), and they are seen rather frequently: 0.5--6 CME/day are detected with typical speed of 250--500\,km\,s$^{-1}$ depending on phase of the solar activity cycle \citep{2010ASSP...19..289G}.

On other stars, however, while flares can be relatively easily observed by photometry, CMEs are harder to detect: they can be recognized by their Doppler signature seen mainly in Balmer-lines. 
The ejected material appears as a blue-wing enhancement of the line, or, in the case of faster events, they could appear as a separate emission bump (or absorption, if seen against the stellar disk). Up to now there are only a handful of CMEs observed, all on dMe-type stars.
A detailed analysis of the observational possibilities and constraints was discussed by \cite{petra-PhD}.

The fastest known event was detected on AD~Leo \citep{1990A&A...238..249H} with a maximum projected velocity of $\approx 5800$\,\kms .
Further events were found on AT~Mic (dMe) by \cite{1994A&A...285..489G} who interpreted these as coronal evaporations, 
on a T-Tauri star \citep{1997A&A...321..803G} 
and on DENIS~104814.7-395606.1, an old dM star \citep{2004A&A...420.1079F}.
Recently \cite{v374peg} analyzed a complex CME event on V374~Peg in good temporal and wavelength resolution with multiple failed eruptions and one eruption that has the maximum projected speed larger than the escape velocity. Until date, this is the stellar CME event that was observed and studied in highest detail. Beside, the event presented by \cite{1997A&A...321..803G} also included a detailed observation of such an eruption.
Stellar CME events in the (F)UV were observed 
on V471~Tau (two events found during 6.8 hours by \citealt{2001ApJ...560..919B})  
and on AD~Leo \citep{2011A&A...536A..62L}.
\cite{2018arXiv180110372F} studied line asymmetries on 28 M-dwarfs in 473 spectra in \HA{} Na {\sc i} D and He {\sc i} lines -- this was the largest such survey to date.

Most of these detections in the literature are single CME events found serendipitously, and are too sparse for statistical analysis. A good strategy could be to obtain observations of several targets in open clusters by multi-object spectroscopy, however, these efforts have not resulted in clear detections yet  \citep{blanco1, iaus}.

Another strategy for searching CME events is to gather all available observations in public archives of a possibly interesting target list.
The first efforts in this search were presented in \cite{iaus}. In this paper we present an analysis with an extended target and archive set.

\section{Data}   
\begin{figure*}
\centering
\includegraphics[width=0.9\textwidth]{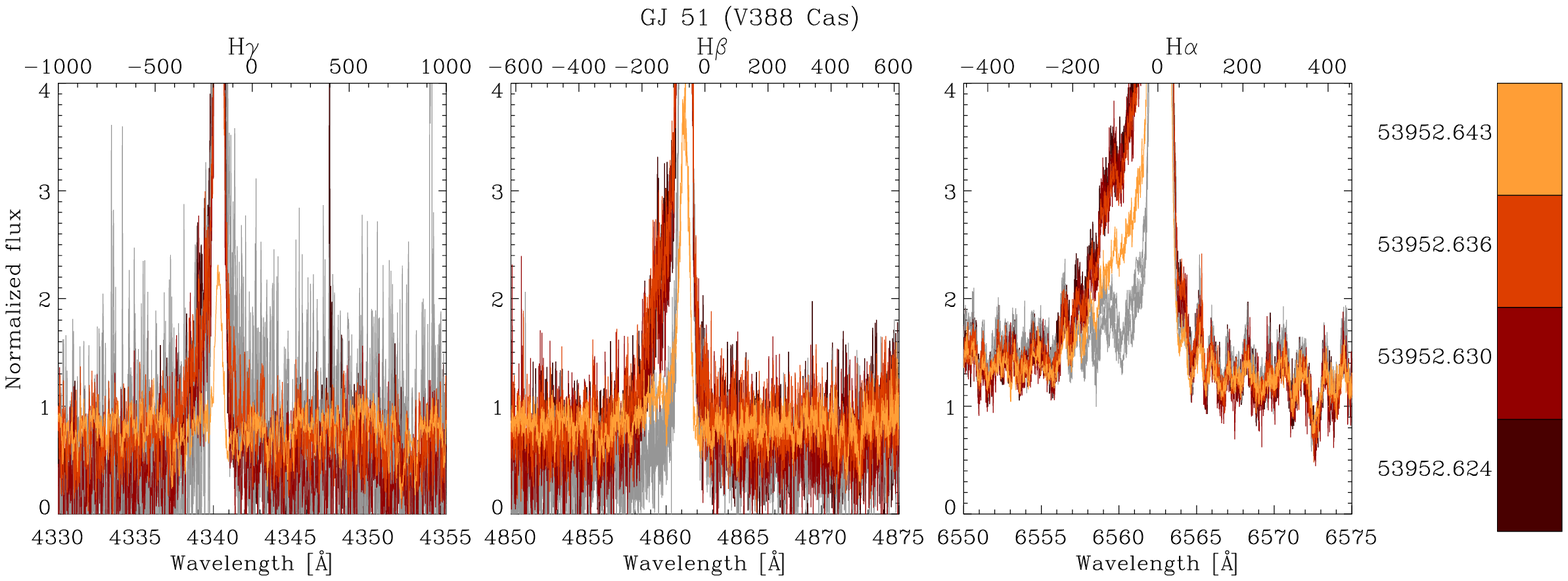}
\includegraphics[width=0.9\textwidth]{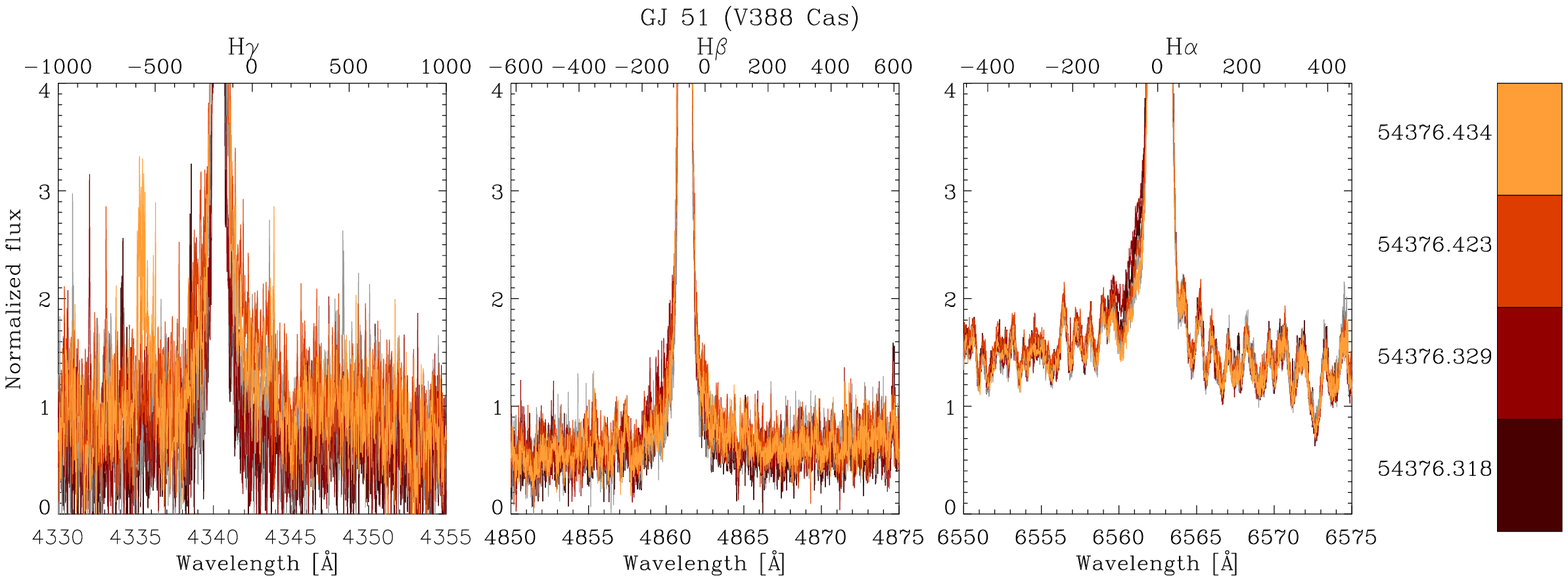}
\includegraphics[width=0.9\textwidth]{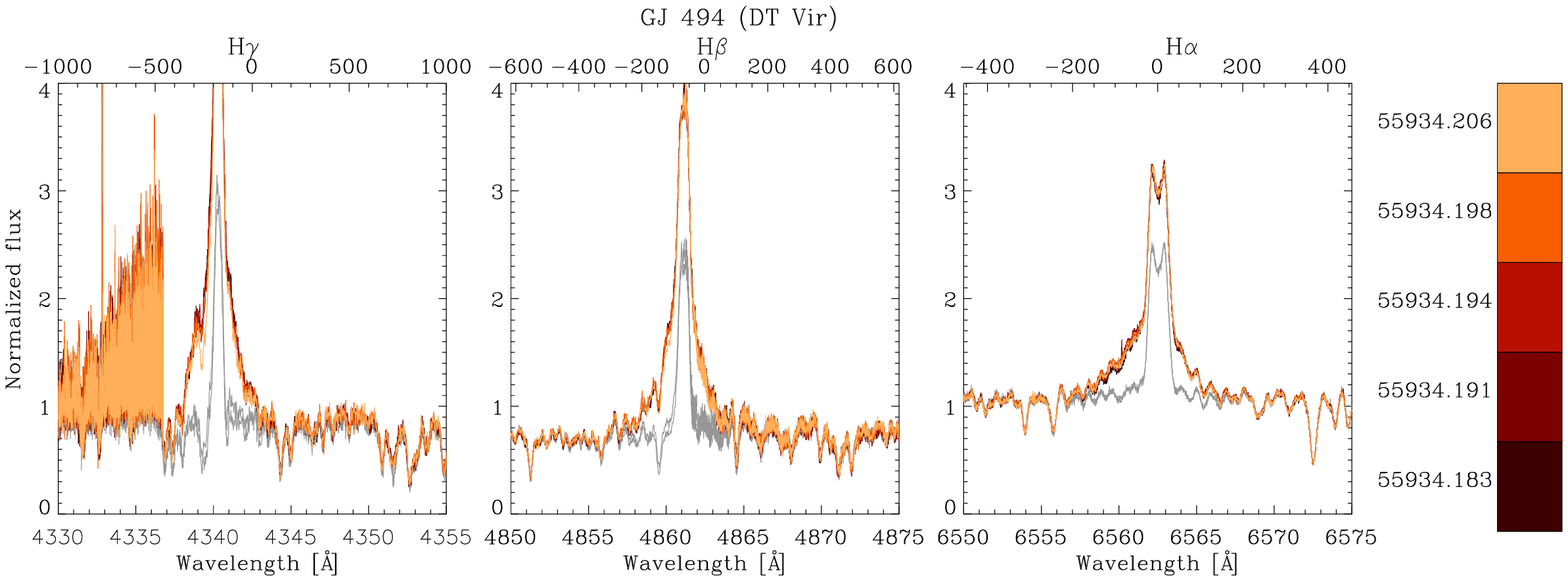}
\caption{Notable events in the Balmer-lines. Bottom plots show the same spectra zoomed in. Spectra outside the event are shown in grey for comparison.}
\label{fig:cmeplots1}
\end{figure*}

For our search we used the list of single late-type stars -- M-dwarfs -- within 15\,pc of the Sun from \cite{petra-PhD}, that includes 382 objects, with some additional prominent objects (G--K--M spectral type).  We focused only on single stars (and wide binaries) mainly because the Doppler-signatures related to CMEs are hard to detect and the effects of binarity might mask them, but also to keep the sample homogeneous. 

We originally started the archive search on the Canada--France--Hawaii Telescope (CFHT) archive site, but we decided to extend and automatize the search using the Virtual Observatory.
The bulk of the analyzed data was downloaded from Virtual Observatory (VO) archives. Since the currently available VO tools are mainly adapted for single objects and a few spectra, we decided to write a \verb+python+ program\footnote{%
available at \url{https://github.com/vidakris/vo.query-spectra} }
to query the VO catalogue for our target list. This program is based on the \verb+astropy+\footnote{\url{http://www.astropy.org}} and \verb+pyvo+ packages\footnote{\url{http://pyvo.readthedocs.io}}. It resolves the coordinates of the given target using the the CDS name resolver, then searches the VO for spectral data using all Simple Spectrum Access (SSA) services within a given radius (we used 5 arcmin for the search). It turned out, that for our purposes only the \verb+Polarbase_SSAP+ (mainly consists of data from the ESPaDOnS instrument on CFHT, but also NARVAL spectra) and the \verb+TBL_Narval+ (data from the NARVAL spectropolarimeter on the Telescope Bernard Lyot) services provided suitable datasets: time-series spectral observations with high resolution and good S/N ratio. Other services in the Virtual Observatory either had just single (or a few) spectra (e.g. \verb+LAMOST.DR1.SSAP+), or did not have enough observations for time series analysis (\verb+SubaruHDS+). After obtaining the data, no additional reduction steps were needed for the further analysis.

The data -- more than 
5500 spectra and 
more than 1200 hours of observation collected -- are summarized in Table \ref{tab:vospec}. 
We note that the exact total observing time was impossible to calculate, as in few cases the \verb+EXPTIME+ (or its equivalent) field was NULL, and in some cases it contained either the year or the month of the observation instead of the exposure time without any other usable information (beginning--end time of the observation, etc.) in the header.

\section{Analysis}   
\begin{figure}
    \centering
    \includegraphics[angle=0,width=0.45\textwidth]{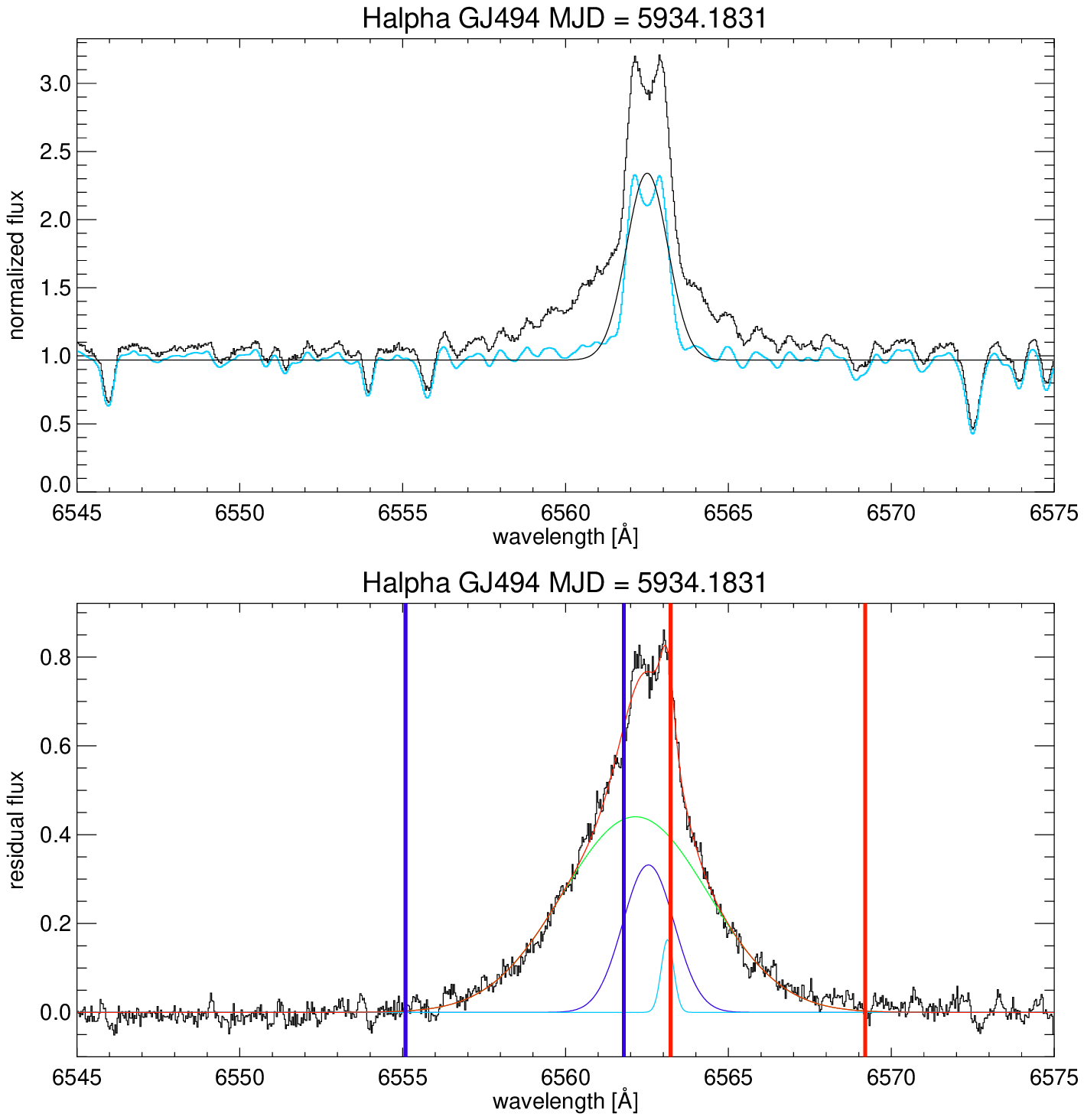}
    \caption{Upper panel: Average (quiescent) spectrum of GJ494 (cyan line) and an example of an active spectrum (black line). Overplotted is the fit (one Gaussian) of the quiescent spectrum. Lower panel: residual spectrum (black line) of the event from the upper panel. Fit of the residual profile (red line) together with the single components (green, blue, and cyan lines). The red and blue vertical solid lines correspond to the integration limits which are used for the determination of the flux of the red and blue asymmetries.}
    \label{fig:residualfit}
\end{figure}
\begin{table*}
    \centering
    \caption{Summary of the detected \HA{} line asymmetries. The number of spectra column indicates those spectra with line asymmetries where the Gaussian fit was done without error (thus, parameters could be estimated), number of event column indicates the number of events spanning  multiple observations. Velocities are showing the determined maximum values (see text for details) in \kms{} units.} 
\begin{tabular}{ccccccccccc}
\hline
\hline
ID &
\textnumero{} of & 
\textnumero{} of & 
$v_\mathrm{blue,min}$ & 
$v_\mathrm{blue,max}$ & 
$v_\mathrm{blue,average}$ & 
$v_\mathrm{red,min}$ & 
$v_\mathrm{red,max}$ & 
$v_\mathrm{red,average}$&
Obs. time&
Events\\
& 
spectra & 
events & 
& 
& 
& 
& 
& 
&
[h] &
per day \\
\hline
GJ    51 &  35 & 10	&140 & 557 & 253 &  89 & 504 & 260	& 21.6  &	11.1	\\
GJ  83.1 &   6 & 2	& 50 & 211 & 125 &  78 & 224 & 132	&  5.8  &	8.3     \\
GJ   170 &   8 & 2	& 60 & 174 & 128 & 121 & 217 & 185	&  6.0  &	8.0     \\
GJ   285 &  56 & 19	& 74 & 635 & 206 &  87 & 532 & 209	& 81.8	&	5.6		\\
GJ   388 &  21 & 9	& 73 & 195 & 145 &  68 & 269 & 169	& 44.2	&	4.9		\\
GJ   406 &   7 & 2	&169 & 361 & 271 & 158 & 274 & 208	&  3.6  &	13.2    \\
GJ   431 &   9 & 3	& 91 & 382 & 224 & 166 & 320 & 242	&  3.7  &	19.6    \\
GJ 493.1 &   6 & 2	& 42 & 228 & 128 &  86 & 246 & 151	&  3.3  &	14.4    \\
GJ   494 &  28 & 12	& 64 & 496 & 208 &  73 & 358 & 198	& 73.5	&	3.9		\\
GJ   729 &  26 & 5	& 43 & 205 &  87 &  47 & 232 & 108	& 48.5	&	2.5		\\
GJ   803 &  11 & 2	& 48 & 288 & 128 &  77 & 322 & 134	& 12.5  &	3.8     \\
GJ   873 &  52 & 11	& 49 & 513 & 195 &  50 & 509 & 204	& 62.4	&	4.2		\\
GJ   896 &   9 & 2	& 37 & 248 & 155 & 177 & 277 & 223	& 17.8  &	2.8		\\
GJ  1111 &   6 & 7	&115 & 291 & 171 &  82 & 268 & 223	& 12.7  &	13.2	\\
GJ  1154 &   2 & 1	&134 & 181 & 158 & 157 & 197 & 177	&  4.5  &	5.4		\\
GJ  1156 &   7 & 4	& 81 & 302 & 185 & 166 & 301 & 230	& 15.5  &	6.2		\\
GJ  1224 &   7 & 4	&132 & 578 & 237 & 142 & 528 & 231	& 11.5  &	8.3		\\
GJ  1243 &   8 & 1	& 97 & 285 & 194 & 208 & 285 & 242	&  2.8  &	8.6		\\
GJ  1245 &   7 & 5	& 94 & 349 & 187 &  85 & 327 & 184	& 21.1  &	5.7		\\
GJ  3647 &   8 & 2	& 94 & 210 & 143 & 125 & 221 & 165	&  4.5  &	10.6	\\
GJ  3971 &   4 & 1	&105 & 226 & 170 & 215 & 302 & 263	&  6.0  &	4.0		\\
GJ  4053 &   3 & 1 	&118 & 219 & 156 & 156 & 256 & 197	&  2.7	&	9.0		\\
GJ  4071 &   5 & 2	& 99 & 207 & 148 & 136 & 232 & 172	&  4.0  &	12.0    \\
GJ  4247 & 108 & 12 & 67 & 791 & 266 &  73 & 508 & 282	& 39.4  &	7.3 	\\
GJ  9520 &  38 & 5	& 40 & 319 & 137 &  45 & 317 & 137	& 99.2	&	1.2		\\
\hline                                              
\end{tabular}                                    

    \label{tab:summary}
\end{table*}

We created plots for each spectrum of the \HA, \HB{} and \HG{} regions for our targets, and after a visual inspection we marked those spectra that showed a spectral asymmetry. We note that in many cases the \HG{} region was seriously contaminated by noise, making this part of the spectrum useless.

Our visual inspection revealed altogether 478 spectra with line asymmetries on 25 objects (see Appendix \ref{sect:notes} for notes on individual objects).
Nine larger events with strong asymmetries were found (additionally to the one analyzed by \citealt{v374peg}) on GJ\,51 (V388\,Cas), GJ\,494 (DT\,Vir) and GJ\,285 (YZ\,CMi). The Balmer-regions of these spectra  are plotted in Figures \ref{fig:cmeplots1}, \ref{fig:cmeplots2}, and \ref{fig:cmeplots3}.
We also derived the ratio of spectra that showed line asymmetries, i.e., the event rates for the different objects. This number indicates the chance that at a given time the object is showing an asymmetric line profile. 

We checked, if the wing enhancements could be results of increased \HA{} emission (e.g. during a flare) by scaling up the quiescent spectra and comparing them to the enhanced ones. We found that the scaled-up spectra do not reproduce well the spectral line wings, i.e., that the detected wing enhancements are real features and not the scaled-up versions of previously unseen asymmetry of the quiescent spectra.

We note, that for the study we used only those spectra that showed asymmetric \HA{} lines. Those measurements that had increased \HA{} emission, but no asymmetry (probably corresponding to a flare event) were not selected for analysis, as the goal of the study was a search for stellar CMEs.

\subsection{Estimating velocities and masses }
\label{sect:cme-fits}

To determine velocities and net fluxes of the \HA{} asymmetries it is necessary to know the quiescent state of the stars. As all of the target stars are active stars, we excluded obvious active spectra, i.e., spectra which show peak and/or wing enhancements, in determining an average spectrum of each star. To estimate velocities and net flux we use the common procedure of building residual spectra (see e.g. \citealt{2018arXiv180110372F}) by simply subtracting the average spectrum from the ''active`` spectra.

To determine velocities we fit the residual profiles with three Gaussian functions to account for blue and red wing asymmetries as well as peak variations. For the majority of active spectra it was not possible to identify bulk velocities as nearly all active spectra show both, blue- and red-wing enhancements at the same time. Therefore we decided to determine maximum velocities only. We define a maximum line-of-sight velocity as the point where the residual profile merges with the continuum. According to our definition this is the case where the residual profile lies 5\% above the continuum. For the majority of residual spectra the usage of three Gaussians is sufficient to reproduce the residual profiles. Some of the profiles show a more complex shape and more than three Gaussians might be necessary to reproduce these complex profiles. The deduced maximum velocities of the complex profiles might represent therefore slight over- and/or under-estimations.

For the determination of masses related to the blue and red flux enhancements we use also the residual profiles and need to set integration limits as we need the fluxes of the blue and red asymmetries without the H$\alpha$ line core to calculate their corresponding masses. Therefore we use the full width at half-maximum (FWHM) of the quiescent Gaussian which accounts for the line core as inner limits and the maximum velocities as outer limits (see Fig. \ref{fig:residualfit}). 
As the CFHT and Narval spectra are provided as normalized spectra we need to add the continuum flux level for each star. As we do not have flux calibrated spectra of the target stars we use instead the relation of \cite{Gizis2002} which connects H$\alpha$ continuum flux and Cousins $R$ magnitude.
The integration of the blue and red asymmetries using the limits given above yields then the fluxes necessary to calculate their corresponding masses. To do so we use the relation from \cite{1990A&A...238..249H} which basically connects the flux of the asymmetry to the number of emitting/absorbing atoms i.e. mass. We caution here that this relation gives order-of-magnitude estimations only. Any more accurate calculation of the mass related to blue and red wing enhancements is only possible by NLTE modelling.
The result of the line asymmetry fits are summarized in Table \ref{tab:summary}.

We also note, that in spatially resolved solar observations  CMEs exhibit a variety of forms, most having the "classical three-part" structure, i.e., 
made up of a core, cavity and leading edge \citep{1985JGR....90..275I}. In many solar cases the core is a filament. Filaments are very prominent on the Sun in \HA , in other stars we can probably observe only this part of a CME event. In this case, however, the mass estimated from \HA{} will be lower than the total mass of the ejecta.

\section{Statistical analysis of  the events}   
\begin{figure}
    \centering
    \includegraphics[angle=-90,width=0.45\textwidth]{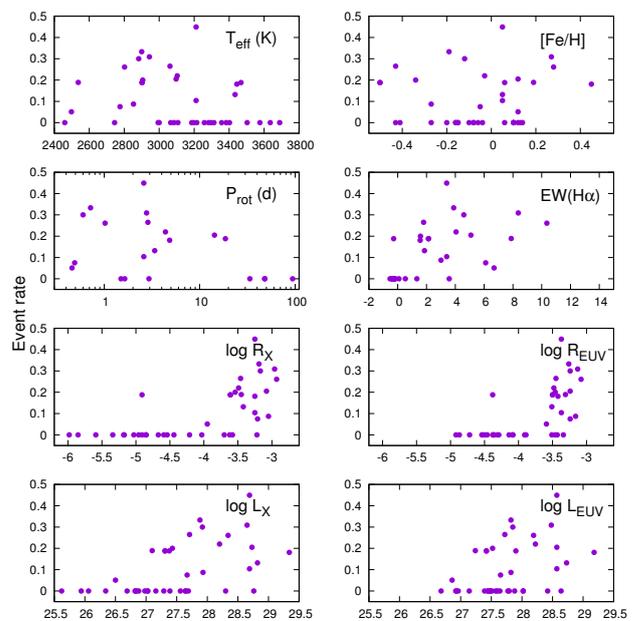}
    \caption{Line asymmetry  rates vs. different physical parameters. EW(H$\alpha$), $\log R_{\mathrm{X}}$ and $\log R_{\mathrm{EUV}}$ denote H$\alpha$ equivalent width (in Angstroms), and X-ray/EUV activity indices, respectively. $\log L_{\mathrm{X}}$ and $\log L_\mathrm{EUV}$ plots are in erg\,s$^{-1}$ units. 
    Note, that for some objects, some information is missing (e.g. $P_\mathrm{rot}$), thus not all subplots contain the same number of points.}
    \label{fig:singleplots}
\end{figure}
\begin{figure}
    \centering
    \includegraphics[width=0.5\textwidth]{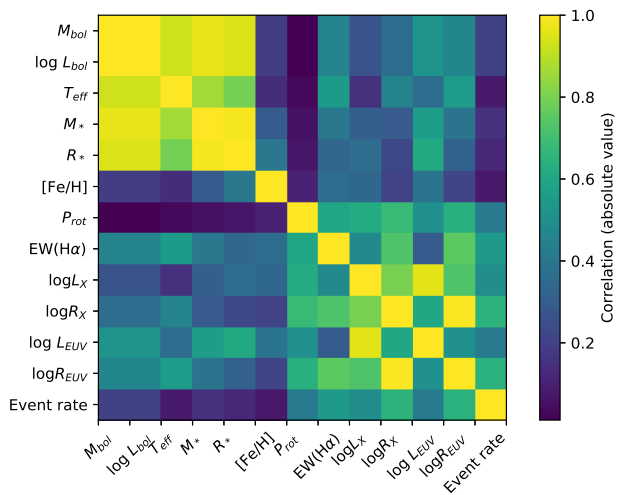}
    \caption{Correlation matrix of the physical parameters. The shade of each box represents the absolute value of the correlation coefficient between the parameters.}
    \label{fig:corr_matrix}
\end{figure}
\begin{figure}
    \centering
    \includegraphics[width=0.5\textwidth]{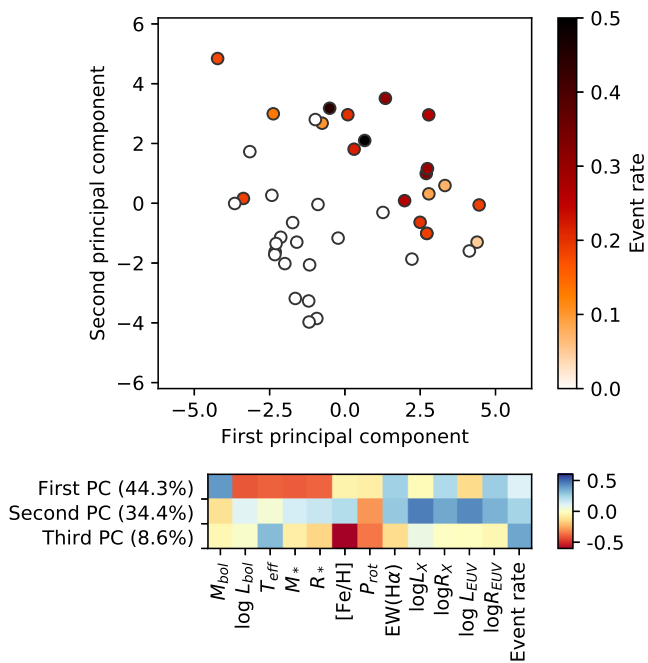}
    \caption{Result of the principal component analysis (PCA). The top plot shows the data plotted along the first and second principal components. Objects with detected line asymmetries  are plotted with filled circles, their shade corresponds to the measured event rate. The bottom plot shows the coefficients for the first three principal components (PCs). The main features of the first PC are related mainly to stellar structure, while the features in the second PC are mainly related to EUV and X-ray activity indices.}
    \label{fig:pca}
\end{figure}

To find a possible relation between the rate of detected line asymmetries and the physical properties of the studied stars, we first plotted the line asymmetry rate versus the physical parameters of the stars (from the appendix of \citealt{petra-PhD}) for the different objects. Here, as event rate we used the rate of those spectra where we detected line asymmetries. 
From this analysis we dropped those targets, where most of the physical parameters were unknown, or there were less than 10 spectra were measured, as in this case distinguishing a distorted line profile from a quiescent one would be uncertain. These plots (shown in Fig. \ref{fig:singleplots}) indicate no obvious relation in case of the $T_\mathrm{eff}$, or the metallicity, but a weak trend might be recognized in the H$\alpha$ equivalent width plot, i.e., stars having larger H$\alpha$ equivalent width have slightly higher  event rates. The measures of X-ray/EUV activity (especially the activity indices) suggest that the event rate increases after crossing a threshold. 
This increase of line asymmetries coincides with the saturation of the activity indices around -3.5---3.0 (see also \citealt{saturation1,saturation2}). 

In Fig. \ref{fig:corr_matrix} we plotted the correlation matrix of the different physical parameters. This suggests that the  event rate is slightly correlated to the   X-ray/EUV activity index and the \HA{} equivalent width (with Pearson correlation coefficients of $r=0.65$, 0.64, and 0.54, respectively). The correlation with X-ray/EUV luminosity ($r=0.50$ and 0.42) and the anti-correlation with the rotation period ($r=-0.42$) are less significant. 

To find further trends/clustering in the data, we also performed principal component analysis (PCA). The purpose of this method is to find a set of linearly uncorrelated variables -- so-called principal components -- from the possibly correlated physical parameters in order to lower the parameter space. 
The transformation is defined in a way that the first principal component has the largest variability in the data, and each succeeding component has the highest variance possible under the constraint that it is orthogonal to the preceding components. 
Intuitively, this can be understood as an $n$ dimensional dataset is rotated in space until we find an $n-1$ dimensional coordinate system where most of the variance in the data can still be seen, then we "collapse" -- reduce the dimensions of the dataset. PCA is often used e.g. in data visualization to get some insights of the available data, or in machine learning, where dimensionality reduction can result in much faster data processing.
The principal components often have no physical meaning, but they can help discovering trends and can also make analysis easier by reducing dimensions. For more details on PCA, see \cite{scikit-learn, scikit-book}.

We found that the changes in physical parameters can be described in 79\% by two, or in 87\% by three principal components. The data plotted along the first two principal components are shown in Fig. \ref{fig:pca}. In this parameter space, the stars showing line asymmetries  seem to form a cluster. According to the PCA, the main features (i.e., physical parameters having the highest contribution) of the first principal components are the parameters describing the stellar structure ($M_*, R_*, T_\mathrm{eff}, M_\mathrm{bol}, L_\mathrm{bol}$) -- these are obviously correlated with each other. The main features in the second principal components were  the parameters describing X-ray and EUV activity (luminosity and activity indices), and -- to somewhat less extent -- the $P_\mathrm{rot}$ and \HA{} equivalent width. Figure \ref{fig:pca} suggests that fast-rotating late-type stars, and objects with high X-ray and EUV activity are the objects that possibly host the events. 
From this plot we can also see, that the inverse correlation of the effective temperature (and correlated values) plays somewhat higher role in the occurrence rate than X-ray and EUV activity, as the objects with detected line asymmetries are better separated from the sample without these along the first principal component.

This means that line asymmetries seem to be more frequent on later-type, more active objects. While this result seems intuitively obvious -- these stars are known to have more flares -- numerical simulations of \cite{cme-suppression} could suggest that strong magnetic fields might be able to block the movement of the material in stellar coronae, if the asymmetries are caused by mainly by CMEs. 

\section{Discussion}   

During the analysis, we were looking for mostly blue-wing enhancements (as we were searching for CME signatures), that would account to material ejected towards the observer, but in many cases red-wing enhancements (material moving away from the observer), and symmetric profile changes were seen. These latter events could be a result of either flares, or -- in the case of a much wider line profile -- expanding material, that have a Doppler-signature in both the  red and blue sides, similar to the light bulb-shaped CMEs seen on the Sun, or a CME occurring near the stellar limb. A further explanation for both blue and red wing enhancements could be backflowing material: on the Sun, this could reach 30--60\% of the total mass of a CME event 
\citep{cmefallback-skylab,cmefallback1,cmefallback2, cmefallback-sim}.
See Sect. \ref{sect:alt} for more details.

In those stars, where line asymmetries were detected, the event rate (with events consisting of multiple spectra) per day was between 1.2--19.6, with an average daily rate of 7.8 event/day (see Table \ref{tab:summary}). In the case of the Sun the typical CME rate is between 0.5--6 per day depending on the phase of the activity cycle.
\cite{v374peg} estimated that the CME rate on V374\,Peg should be in the order of 15--60 CMEs per day. The results from this study are below these values. 
This could have multiple reasons: projection effects or possibly magnetic suppression of the events (see \citealt{Drake, cme-suppression}). 
We note, that comparisons with the solar observations should be handled with care -- first, the methods of observations (the methodology, temporal and spatial resolution) are different in the solar and the stellar case. The sensitivity is also different: most solar CMEs would be impossible to detect on other stars. Also, the Sun is much less active than the  stars studied in this paper, and simply scaling up the solar case could yield an incorrect estimation. 

\subsection{Velocity and mass distribution}
\label{sect:velocities}
\begin{figure}
    \centering
    \includegraphics[angle=-90,width=0.45\textwidth]{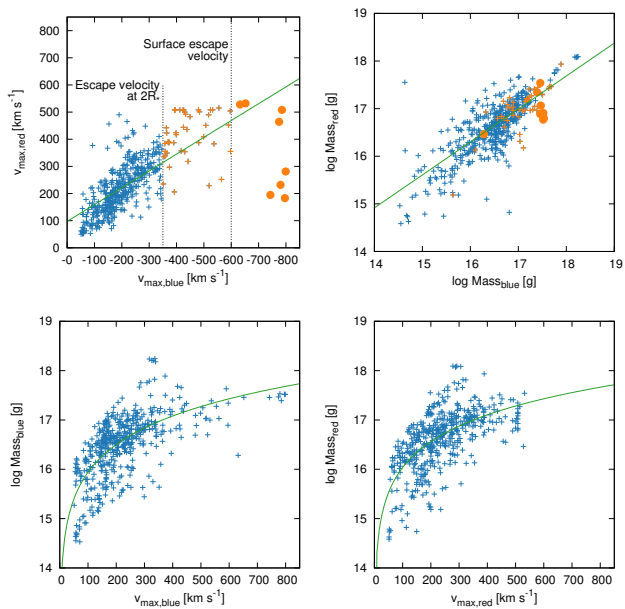}
    \caption{Top row: Velocities and masses derived from blue vs. red enhancements. On the left plot, vertical lines mark the approximate escape velocity on M dwarfs at the surface and at 2 stellar radii, events with higher speed are marked with different colours on both plots. Lines show linear fits to the data.
    Bottom row: relation between maximum velocities and estimated masses.}
    \label{fig:fits}
\end{figure}
\begin{figure}
    \centering
    \includegraphics[angle=0,width=0.45\textwidth]{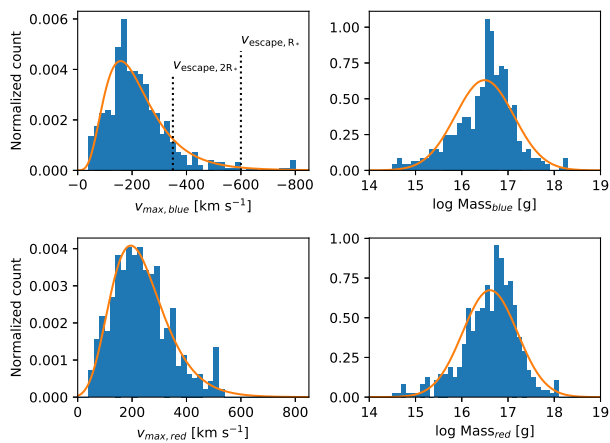}
    \caption{Distribution of CME velocities and masses from blue and red enhancement. Dashed lines show the approximate escape velocities at the surface and at two stellar radii.}
    \label{fig:hist}
\end{figure}

The distribution of velocities and masses  from our study (see Sect. \ref{sect:cme-fits}) are plotted in Figs. \ref{fig:fits} and \ref{fig:hist}. Note, that the velocities from this study yield maximum line-of-sight velocities of the events, not bulk velocities that represent the main velocity component.
From these plots we can see that most of the detected events do not reach surface escape velocity (which is roughly 600\,\kms{} on M dwarfs). Furthermore, it is also worth to note, that the mass and velocity distribution of the blue and red enhancements, e.g. rising and falling back material is very similar. Linear fit to red vs. blue velocities (see Fig. \ref{fig:fits}) yield 
\begin{equation}
v_\mathrm{max,red} = 0.62\pm 0.03 v_\mathrm{max,blue}+97.3\pm 7.0,
\end{equation}
while a similar fit to the logarithm of minimum mass in blue vs. red gives
\begin{equation}
\log M_\mathrm{red} = 0.69\pm0.03 \log M_\mathrm{blue}+5.2\pm 0.4.
\end{equation}
This indicates, that -- if these are connected to ejected/falling-back material -- about 60--70\% of the ejecta either falls back to the surface, or has parts that moving away from us during its expansion (e.g. in the case of an event observed at the stellar limb).
Such red enhancements are not unexpected, as these can be observed on the Sun, where -- depending on the event -- the mass falling back can reach up to 30--60\% of the total CME mass \citep{cmefallback-skylab}. 

We note, that -- if we interpret the events  as CMEs -- the ejecta travel typically $0.1-1R_\odot$, i.e., approximately $1-2R_*$ during the observations. At these distances, the escape velocity is lower than on the surface (roughly 350--440\,\kms ), however, CME acceleration can act beyond these distances on the Sun \citep{cmefallback1}. 
Since -- at least on the Sun -- the ejecta are often accelerated while near the surface, it is very likely, that a larger fraction of the events could be "successful" ones (see Fig. \ref{fig:fits}): roughly 11\% (with projected maximum velocities higher than 350\,\kms ) instead of 1.7\% (with projected maximum velocities higher than 600\,\kms ).

We checked a possible relation between the estimated masses of the ejecta and their velocities (see bottom row in Fig. \ref{fig:fits}). We found that these seem to follow a power-law like relation:
\begin{equation}
\log M = a v^k.
\end{equation}
The best fitting values for $a$ and $k$ were 
$a_\mathrm{blue}=12.67 \pm 0.17$, 
$k_\mathrm{blue}=0.050 \pm 0.003$ and
$a_\mathrm{red}=12.95 \pm 0.18$, 
$k_\mathrm{red}=0.046 \pm 0.003$ for the blue and red enhancements, respectively. 

Majority of the events are correlated with an enhanced peak of the Balmer-lines. This would mean that these events are almost always related to a flare, similarly to the case of the Sun. 
Moreover, most enhancements are very symmetric and only a very few cases show a distinct blue wing enhancement (a signature of material moving into our direction). 
However, these facts do neither rule out nor confirm the possible alternative explanations (see Sect. \ref{sect:alt}). Correlation with flares could also mean the event is chromospheric evaporation/condensations or line broadening.

We found that the  typical observed maximum velocities of the events are in the order of 100--300\,\kms{} (see Fig. \ref{fig:hist}). 
Typical solar chromospheric evaporations have velocities in the order of several tens of \kms{} -- sometimes reaching a few hundred \kms{} in hot emission lines , while solar CMEs have velocities in the order of 250--500\,\kms{} (up to $>2500$\kms , see \citealt{cme-livingreview}).
A lognormal distibution fit to the blue and red maximum velocities yield 
$\mu_\mathrm{blue}=5.41, \sigma _\mathrm{blue}=0.45$, 
$\mu_\mathrm{red}=5.91, \sigma _\mathrm{red}=0.29$ and a peak of 
$v_\mathrm{blue}=$200\,\kms{} and $v_\mathrm{red}=$222\,\kms{}, respectively.
The typical masses  are in the order of $10^{15}-10^{18}$g. 
A normal distibution fit to the blue and red masses (note: the masses are in logarithmic units) yield 
$\sigma _\mathrm{blue}=0.63$, 
$\sigma _\mathrm{red}=0.59$ and a peak of 
$\log M_\mathrm{blue}=$16.49\,g and $\log M_\mathrm{red}=$16.61\,g, respectively.

The detected maximum velocities are often lower than the escape velocity on the studied objects (see Fig. \ref{fig:fits}), which could have multiple reasons. 
{\emph a)} We just see projected velocities, the true velocities are higher, and more CMEs are leaving the surface. This projection effect could be worsened by an unintended selection effect in the sample: e.g. if a large fraction of the targets were observed with the intention of Zeeman--Doppler mapping or Doppler imaging, they could have very similar inclinations, meaning that the inclination distribution of our sample is not random (cf. the discussion on comparison with ZDI maps in Sect. \ref{sect:zdi}). The ratio of red/blue masses disfavors this option, however, we do not know if these stars have preferred source locations. 

{\emph b)} We can observe only the early phase of the events (cf. Fig. 5 in \citealt{cmefallback-skylab}). 
On the Sun only the fastest CMEs reach escape velocity near the solar surface.
In most cases, the observed solar CMEs do not have a constant velocity, they are often accelerated near the Sun (they are mainly accelerated in the lower corona, within $2R_\odot$), reaching escape velocities only at higher altitudes. 
The slowest events tend to show acceleration while the fastest tend to decelerate in higher regions (see \citealt{cme-livingreview} and references therein).
However, as the material is moving, the outside pressure in the atmosphere is decreasing. This yields to a (partly adiabatic) expansion of the ejecta, resulting in cooling of the material with its density decreasing. 
If the expansion and cooling of the material is fast enough, it is possible, that we can just observe the very first phase of the CMEs, while their acceleration is still in progress. If the acceleration continues to large distances, then the fastest CME signature cannot be detected in the Balmer lines any more due to the density decrease of the material.
On the other hand, solar CMEs have been observed in \HA{} up to several solar radii -- before the emission switches to Thomson scattering (scattering of light from free electrons)
due to gradual photoionization \citep{Howard2015a,Howard2015b} -- using coronagraphs with \HA{} filter
(see e.g.
\citealt{Sheeley1980, House1981, Dryer1982, 1985JGR....90..275I, Mierla2011, Howard2015a} ), 
although the coronal structure of M-dwarfs could be somewhat different from that of the Sun -- in hotter coronae prominences would get ionized earlier (i.e., at smaller distances). This scenario would not explain all the observations per se, as strong events -- like the one on V374\,Peg \citep{v374peg} or the CME reaching 5800\,\kms{} on AD\,Leo \citep{1990A&A...238..249H} -- still can be seen in Balmer lines.
It could be possible, that these stars have huge cool eruptive prominences, similarly to the 2011 June 7 event observed on the Sun \citep{solarprominence1,solarprominence2,solarprominence3}.
Such large prominences have also been observed on other stars, e.g. on EY Dra \citep{1998A&A...337..757E}, on HK Aqr and on PZ Tel \citep{2016MNRAS.463..965L}.

{\emph c)} These findings indeed describe the velocity/mass distribution of the CMEs. This would mean that on M-dwarves, successful CME events are sparse, and only a handful CMEs are actually leaving the surface. Based on numerical models, \cite{Drake} and \cite{cme-suppression} suggested that magnetic suppression -- i.e, that strong magnetic fields would prevent material leaving the stellar surface -- could be a viable mechanism to weaken CMEs on very active stars. This would cause, that weaker events would be suppressed and only "monster" CMEs could build up enough energy and speed to be able to break free, and only these events would behave as the ones we see on the Sun.

It is likely, that the real scenario is a combination of all the above. True maximum velocities are probably somewhat higher due to projection effect (although bulk velocities will be slower than the values reported here). In the case of a typical ZDI target with an inclination of $60\degree$, supposing a CME from the equatorial region (as mainly seen in the case of the quiet Sun, that has a dipole-like magnetic field, similar to M-dwarfs) would cause a difference in the order of $\approx 5-20\%$. The observed ejections could fall into two categories. Weaker events could be suppressed by the magnetic field, or diluted/ionized by the time they reach escape velocity while accelerating, making them unobservable in the \HA{} regime. Only the strongest CMEs would start already with high velocity, and thus be detectable in \HA{} in the early stages of the eruption.

\cite{2018arXiv180110372F} studied  473 spectra of 28 emission-line M dwarfs to search for line asymmetries. They found 63 such observations, and concluded that only 4\% of \HA{} asymmetries are connected to similar variations in the Na {\sc i} D and He {\sc i} D$_3$ lines.
In their survey the authors found only weak enhancements -- none of the detected asymmetries were beyond 6560\AA{} -- confirming our results, that most of the events are weaker ones. They explained blue wing enhancements by chromospheric evaporation, and red wing asymmetries by coronal rain/chromospheric condensation.

\subsection{Possible explanations for line asymmetries}
\label{sect:alt}
Asymmetries in the Balmer-lines -- especially in the \HA{} region -- are generally associated with moving material in the chromospheres: red asymmetries with downward motions (also known as coronal rain or chromospheric condensation) or during flares cooling flows along (post)flare loops can also contribute to redshifted \HA{} profiles. Blue asymmetries could be related with ejected material (e.g. CMEs) or chromospheric evaporation \citep[see e.g.][]{2018arXiv180110372F}.  
This latter suggestion  -- chromospheric evaporation -- would explain why the blue-shifts have such low typical velocities (for more detail see Sect. \ref{sect:velocities}). 

Typical chromospheric evaporation velocities in solar flares are several tens of \kms , but explosive chromospheric evaporation velocities can reach velocities in the low hundreds. However, these velocities are observed only in hot coronal lines (e.g. in Fe {\sc xix}), while in the cooler lines (He {\sc i}, O {\sc v}, Mg {\sc x}) the observed flows over flare ribbons are redshifted downflows of 20--50 \kms  (see e.g. \citealt{evaporation-speed1, evaporation-speed2}). The spectra we analyse here cover the cool Balmer lines, where flows above the flare ribbons should mainly be redshifted. Indeed, in the \HA{} line, \cite{2012PASJ...64..128A} find strong red asymmetry corresponding to $\approx$50 \kms{} \textit{downflows} over the chromospheric flare ribbons in an X2.3 flare. Although \cite{evaporation-sim} find upflows in the Lyman lines in their simulations of the flaring atmosphere using the RADYN code, the upflow velocities they find reach only a few tens of \kms .
Based on simultaneous \HA{} and X-ray observations \cite{1990ApJ...363..318C} reported rare blue-wing enhancements (possibly due to chromospheric heating) with velocity in the order of 100\,\kms{} embedded in a generally redshifted plasma-motion environment during the impulsive phase of solar X-ray flares, where some kind of ejecta was likely involved (probably connected to the heated part of an erupting filament or a jet). 
On AT Microscopii \cite{1994A&A...285..489G} considered an event connected to a flare, that reached $\approx600$\,\kms{}  maximal velocity in the Ca {\sc ii} H\&K and Balmer lines, what was interpreted as due to "high-velocity evaporation". 
This may indicate that flaring conditions on dMe stars are significantly different from those on the Sun, or that the observed strongly blueshifted flows were due to a CME instead.

The typical average CME speeds on the Sun are 250--500\,\kms , depending on the phase of the activity cycle, but the  apparent speeds of the leading edges of CMEs range from about 20 to more than 2500\,\kms{} \citep{cme-livingreview}. 
We note, that these velocities are measured in white light, quite high up in the corona. In \HA{} we would only observe low velocities of the accelerating CME very close to the Sun.
Unfortunately, from the maximal projected velocities only -- that we can measure from the spectra -- we cannot be certain about the exact nature of the observed phenomena.
Direct comparison of solar and stellar observations is further hindered by the fact that solar spectrographs have typically no broad wavelength coverage:they cover only the \HA{} profile itself but not the blue/red continuum.
Therefore there are few papers in the literature that allow direct comparison between solar and stellar data.
\cite{1993ARep...37...76D} presented
\HA{} filtergrams and spectrograms of an M7.3 solar flare on 1989 March 12. Strong ejection of material was detected with velocities up to 300--600 \kms . Their Fig. 1 show \HA{} line asymmetries similar to the stellar events analyzed in our paper.
During an M2.6 solar flare on 2002 September 29 \cite{2003ApJ...598..683D} observed an erupting filament that reached a line-of-sight velocity of  $\approx210$ \kms{} in \HA{} spectral data. The authors noted that some parts of the filament produce emission in the \HA{} blue wing, i.e. that the filament gets heated during the eruption and emits (instead of absorbs) in \HA .
Multiwavelength Skylab data showed "\HA{} emitting material in mass ejections from flare sites", that was present in 9 out of 10 flares \citep{1979SoPh...61..201M}.
Based on Skylab coronograph data the authors found a correlation of CMEs and chromospheric (\HA ) activity, and correlations of CMEs and eruptive prominences.
Furthermore, statistical analyses of Nobeyama microwave data of erupting filaments indicated that the upward velocities are comparable with stellar data in our paper:
\cite{2002A&A...382..666H} found 50 filament eruptions up to 114 \kms{} upward velocity.  
\cite{2003ApJ...586..562G} studied  filament eruptions and found an average velocity of  $v \approx 65$ \kms{} (based on 147 filaments) -- 34 out of 147 filaments had radial velocity $v> 100$ \kms{}, up to 380 \kms .
In stellar data, we most probably observe erupting filament material, which gets heated and emits in \HA{} during the early phase of the eruption (i.e. before it would get further heated and disappear from the \HA{} line, see also the discussion in Sect. \ref{sect:velocities}b). The few solar papers above indicate that the radial velocities measured in erupting solar filaments are comparable to those in stellar data.

From the existing attempts to detect stellar CMEs it seems, that fast and massive events are rare, and by examining the \HA{} region we also cannot distinguish slower (projected) and less massive events from other \HA{} plasma motions. 
Our analysis seems to confirm these findings. Most of the measured velocities (unless heavily distorted by projection effects) are below the surface escape velocity (see Fig. \ref{fig:fits} and Fig. \ref{fig:hist}), thus cannot be successful CME events. This does not change significantly if we suppose that the ejected material is accelerated near the stellar surface (see discussion in Sect. \ref{sect:velocities}). Thus, we can conclude that if the detected line asymmetries are connected to plasma flows, the moving material is probably not ejected to the circumstellar space in most cases (90--98\%).

However, chromospheric flows are not the only possible explanation for these asymmetries in the line profiles. 
It is also possible, that steep velocity gradients in the flaring chromosphere can cause opacity changes at different wavelengths, which would yield to observable red and blue asymmetries \citep{Kuridze2015}. 

It is also worth mentioning that strong stellar wind could mimic line enhancements of CMEs, but that is probably an unlikely scenario. Depending on the magnetic configuration, inclination, rotational phase, etc., stellar wind could distort the profiles rather asymmetrically and if the phase coverage is sparse, it could be difficult to differentiate between CMEs and stellar wind, although in the case of the Sun, solar wind is too hot and tenuous for detection in Balmer-lines. We note, that on  M dwarfs hot and tenuous stellar winds can be detected via  Lyman\,$\alpha$ absorption \citep{wood-windreview}. However, in the latter case, line distortions of this type should be seen constantly, thus we assumed that the dynamic line enhancements originate from CMEs. 
Only recently, \cite{Pavlenko2017} report on blue-shifted emissions seen in Balmer lines which they interpreted as wind signatures shifted by typically 30\,\kms , which is rather slow. Stellar wind in cool stars is measured as interaction with the interstellar medium as astrospheric absorption seen in Ly$\alpha$ \citep{stellarwind1,stellarwind2},  or as free-free emission originating from fully ionized winds \citep{Gudel2002, Gaidos2000,Fichtinger2017}, and its detection was also attempted through radio observations (see \citealt{stellar_winds, Gudel2002} and references therein).

\subsection{Comparison with Zeeman--Doppler maps}
\label{sect:zdi}
The velocities derived from the spectra are projected, and just from the spectral data it is not possible to know their origin and thus their actual speed. 
For the stars showing the largest events (shown  in Figs. \ref{fig:cmeplots1}, \ref{fig:cmeplots2} and \ref{fig:cmeplots3}) we checked the literature for available maps of the magnetic field that are relatively close to the events in time in the hope to constrain the source region of the events. 

In several cases Zeeman--Doppler imaging (ZDI) maps from the literature were only available from different epochs, or the phase coverage of the original observations was too poor for a reasonable comparison. For the sake of completeness, these efforts are summarized in Appendix \ref{sect:appendix-zdi} in detail.

There are available ZDI maps close to the detected events in the case of GJ 51, YZ CMi and V374 Peg. In the case of V374 Peg a Doppler map is also available. 
This enables us to investigate the events together with the magnetic field configuration for the first time: there are no examples in the literature where both observations (i.e., magnetic/surface maps and time series spectra of line asymmetries) are available from the same epoch. These data could be crucial for future modelling efforts of such events. 
The observations suggest that all these three objects are quite similar: they all have inclination of 60--70$^{\circ}$, and they all possess an axisymmetric, poloidal magnetic field.
The ZDI maps indicate that the strongest recovered magnetic field strength on GJ 51, YZ CMi and V374 Peg are roughly 4, 3, and 1\,kG, respectively -- these were radial magnetic field strengths. According to the maps, the strength of the azimuthal and meridional field is roughly half of these values. The average magnetic fields on GJ 51, YZ CMi and V374 Peg were 1.6, 0.6 and 0.7\,kG, respectively. These values are much larger than the one used by \cite{cme-suppression}, who assumed a simple 75\,G
dipole aligned with the rotation axis of the star for the numerical simulation.

We considered three scenarios: 
{\em a)} in the first case we suppose that the  events are connected with one of the large active regions. 
{\em b)} In the second case we suppose that the events originate from around the equatorial region (between the two large active nests), as in the case of the quiet Sun, that has an axisymmetric, poloidal magnetic field. 
Most of the solar filaments are located in-between active regions/boundary of active regions (cf. Fig. 24 in \citealt{Parenti2014} and \citealt{2008ApJ...686.1432G} describing a huge filament which formed between two active nests)
Here the CMEs mainly originate from equatorial steamer regions (see \citealt{cme-livingreview} and references therein)  as the result of the interaction between the slow solar wind and the magnetic field. 
However, solar streamer-blowout CMEs are usually not accompanied by flaring.
{\em c)} In the third scenario the CMEs events are originating from the smaller-scale magnetic field -- in this case we can obtain no further information on their origin.
On the Sun, these are the so-called "quiescent prominence" eruptions, which originate from the decayed/dispersed remnant field of a former active region. These can lead to large CMEs, but they are usually not accompanied by flare events, as the magnetic field is too weak and the magnetic reconnection rate is too low for observable flare brightenings.

In the case of V374 Peg the eruptions of the complex CME event were seen at phases 0.72, 0.89 and 0.97 (using the same phasing as \citealt{zdi-v374peg}). According to the Doppler map,  the stellar surface is unspotted between phases 0.70--1.05. There is no evidence of a polar spot either. 
Thus in scenario {\em a)}, if the CMEs are connected to one of the active regions, the three events should be connected to different active nests. 
In this case, the first event (referred as BWE1 in \citealt{v374peg}) is connected either to the active nest at phase 0.70 or the one at 0.63 (both regions are in the plane of sight). 
Here the true velocity of the event is either the measured $-350$\,\kms{} or slightly higher, $\approx -385$\,\kms{}, respectively. 
BWE2 was observed at phase 0.89, where no spots are seen. Thus, this event could be connected to either the active nest at phase 0.70 (in this case the measured $-350$\,\kms{} projected velocity would correspond to $-950$\,\kms), or to the one at phase 0.05 (with a velocity of $\approx 650$\,\kms). 
In this case, BWE2 also reaches escape velocity, which was estimated to be $v_e\approx 580$\,\kms. BWE3 occurred at phase 0.97, when only the active nest at phase 0.05 was in view -- this would mean that the measured projected velocity $v_\mathrm{proj}=675$\,\kms{} would correspond to $v=770$\,\kms{}. 
To sum up, in scenario {\em a)}, if we suppose that the CMEs are connected to the active nests, the three events are probably connected to two different active regions, and BWE1 is still under the escape velocity, but both BWE2 and BWE3 is above it.
In scenario {\em b)} we suppose that the CMEs originate from the equatorial region, as seen on the quiet Sun, that also has an axisymmetric, poloidal field, similar to these objects. Here it is possible that all three events are connected to the same region. 
In this case, the latitude of the CME is constrained to a rather thin range around phase 0.83, so the source of the event can be seen both from the time of BWE1 and BWE3. This would mean that the real speed of the three events are 
$v_\mathrm{BWE1}\approx 455$\,\kms ,  
$v_\mathrm{BWE2}\approx 375$\,\kms , and
$v_\mathrm{BWE3}\approx 1060$\,\kms , respectively. The uncertainty in latitude would add an additional increase in the order of $\approx 10\%$.

On GJ 51 two large events were detected: one in 2007 October and another in 2006 August. As the available ZDI maps have poor phase coverage, a detailed comparison with surface features is not reasonable. 
In both events, the maximum velocities from the fits are approximately 355\,\kms{} (we note that in the case of the 2006 event the fits were flagged as problematic). In the case of scenario {\em b)}, when the eruption is originating from the equatorial region, the measured  velocities increase  by 4--40\% up to $v=370-500$\,\kms ,  supposing that they are originating from an equatorial stripe ranging from $-15\degree$ to $+15\degree$ (with stellar inclination of $i=60\degree{}$, see \citealt{zdi-late}), and that event was observed at the phase of the eruption. These latitudes correspond to the typical CME latitudes that are observed on the quiet Sun. This would mean that these events were both below the escape velocity, thus -- in case of no further acceleration -- are failed eruptions.

The three large events on YZ CMi (shown in Fig. \ref{fig:cmeplots2}) were observed at the same time as the data for the ZDI maps were obtained. They occurred at phases 0.85--0.86, 0.39--0.75, and 0.8 between HJDs 2454486 and 2454508. If these events are connected to the active regions (scenario {\em a)}) the eruptions could be connected either to the strong polar active nest, or in the case of the two shorter events at 0.85 and 0.8 might be also connected to the southern region of negative polarity, but this is less likely, as only a small part of the southern active region is visible. If we suppose that the eruptions are connected to the pole, the measured $v_{1,\mathrm{proj}}=635$\kms{} $v_{2,\mathrm{proj}}=415$\kms{} and $v_{3,\mathrm{proj}}=390$\kms{} would translate with an inclination of $i=60\degree$ \citep{zdi-mid} to $v_1=1270$\kms , $v_2=830$\kms{} and $v_3=780$\kms . In the case of the second, longer event, the peak velocity was reached at rotation phase 0.75, while the event itself occurred while the active region covering the surface from the pole roughly to the equator was visible. If the event happened roughly at the center of this region, this would mean that it was observed with a phase difference of $\approx 0.175$, i.e., $60\degree$ , meaning that the projected velocity is half of the true velocity, yielding $v_2=830$\,\kms .

\subsection{Consequences on the circumstellar environment and exoplanetary atmospheres}

 As mentioned in the introduction, flares and CMEs can have a serious impact on their environment by gradually evaporating planetary atmospheres
\citep{2007AsBio...7..167K,2008SSRv..139..437Y}.
Our findings suggest that mass ejections leaving the star are relatively rare events on late-type active stars.
This would confirm the results of \cite{cme-suppression}, who -- based on a numerical study -- suggested that a large-scale dipolar magnetic field of 75\,G could be able to fully confine eruptions within the stellar corona, and  only the largest eruptions would leave the stars.  
We found that the detected line asymmetries happen on cooler, more active objects, but even here, 90--98\% of the events are detected below escape velocity and could be more likely connected to chromospheric evaporation than CMEs.  
This would suggest that the strong magnetic field of the host star could mitigate CME hazards (similar to the conclusions of \citealt{mullan-trappist1} in the case of the TRAPPIST-1 system) and the more active stars could provide a safer environment for exoplanetary systems than previously thought.
On the longer term, in the star–exoplanet relations, atmosphere loss due to enhanced high energy radiation (e.g.  \citealt{2014MNRAS.439.3225L}), typically found in young stars, and a possible contribution of flares would play a stronger role than CMEs.

\section{Summary}   

   \begin{itemize}
 \item We analyzed spectral data of single stars from telescope archives and the Virtual Observatory database. The Balmer regions were visually investigated for asymmetric wing enhancements that could indicate Doppler signature of ejected material, i.e., possible coronal mass ejections (CMEs);
 \item Of more than 5500 spectra 478 spectra with line asymmetries were found on 25 targets, including nine larger events -- this is the largest survey of this kind to date;
 \item The wing enhancements cannot be reproduced by simply scaling up the quiescent spectra;
 \item The events were modelled using three-component Gaussian curves, based on these the maximum velocities and the masses of the ejecta were estimated;
 \item If we interpret the events as CMEs, we find that most of the detected events (90--98\%) do not reach escape velocity while being observed. The typical maximum velocities and estimated masses are in the order of 100--300\,\kms{} and $10^{15}-10^{18}$\,g, respectively. The masses and velocities of the ejecta seem to be related by a power-law function; 
 \item These estimated velocities could be distorted by projection effects or it is possible that we only see an early phase of the events, while they can still be detected in the Balmer regions, before their acceleration in the higher atmosphere. It is also possible that these events are suppressed by the strong magnetic field of the star;
 \item The detected event rates were in the order of 1.2--19.6 event/day (on the Sun this is 0.5--6 CMEs/day depending on the phase of the activity cycle). These values are still somewhat lower than expected from the solar case (15--60 event per day), but this could be -- at least partly -- explained by observation effects;
 \item In some of the events, Zeeman--Doppler maps were available near the line asymmetries. In such cases we attempted to estimate the true velocities of the events supposing different scenarios for their origin;
 \item A statistical analysis of the event suggests that the occurrence rate of Balmer-line asymmetries is higher in later-type stars that have faster rotation rate, and have stronger chromospheric activity. The events seem to occur only after reaching a threshold in chromospheric activity.
 \item The relatively low typical velocities, the high ratio of falling-back material, and the rarity of strong, fast eruptions could suggest that even later-type, active dwarfs could be a safer environment for exoplanetary systems CME-wise, and atmosphere loss due to radiation effects would play a stronger roles in exoplanetary atmosphere evolution than CMEs.
    \end{itemize}

\begin{acknowledgements}
We would like to thank the referee for their thought-provoking comments, which greatly helped us to re-think several aspects of the work presented in the manuscript, and as a result, improved its clarity.
We acknowledge the financial support of the Austrian--Hungarian Action Foundation (contracts 95öu3 and 98öu5). 
The authors acknowledge the Hungarian National Research, Development and Innovation Office
grants OTKA K-109276, OTKA K-113117, and supports through
the Lend\"ulet-2012 Program (LP2012-31) of the Hungarian Academy of Sciences,
and the ESA PECS Contract No. 4000110889/14/NL/NDe. 
KV is supported by the Bolyai J\'anos Research Scholarship of the Hungarian Academy of Sciences. 
M.L. and P.O. acknowledge the Austrian Science Fund (FWF): P30949-N36.
HK acknowledges the support from the Danish foundation Augustinus fonden.
This research made use of Virtual Observatory data, and the pyVO package, originally developed by the Virtual Astronomical Observatory (VAO).
This research used the facilities of the Canadian Astronomy Data Centre operated by the National Research Council of Canada with the support of the Canadian Space Agency.
\end{acknowledgements}

%
%

\bibliographystyle{aa} 
\bibliography{paper}
\begin{appendix}   
\section{Notes on individual stars/events}
\label{sect:notes}
\begin{description}
\item[GJ 51 (V388 Cas)] Two strong events (shown in Figure \ref{fig:singleplots}). The one on HJD245392 is most pronounced in \HA . Nine further, weaker blue profile enhancements.
\item[GJ 83.1 (TZ Ari)] Two weaker events. The second, broader one is better visible in \HB{} and \HG{}.
\item[GJ 170 (V546 Per)] One event, both the red and blue wings are enhanced.
\item[GJ 285 (YZ CMi)] Several weaker BWEs, and three strong ones (plotted in Fig. \ref{fig:singleplots}). The event on HJD2454487 shows a broad, enhancement, which is somewhat stronger in the blue wing. The HJD2454494 event shows a stronger red wing enhancement.
\item[GJ 388 (AD Leo)] Several slow blue wing enhancements.
\item[GJ 406 (Wolf 359)] Six weaker events.
\item[GJ 431 (V857 Cen)] One stronger blue end red wing enhancement.
\item[GJ 493.1 (FN Vir)] Two events, the one on HJD2456352 shows a symmetrical wing enhancement.
\item[GJ 494 (DT Vir)] Four weaker, one stronger, symmetrical enhancement. One distinct eruption, plotted in Fig. \ref{fig:singleplots}.
\item[GJ 729 (V1216 Sgr)] Continuous change of the H$\alpha$ profile, three smaller BWEs.
\item[GJ 803 (AU Mic)] Only one slow BWE was detected.
\item[GJ 873 (EV Lac)] Five slow, weak BWEs; two stronger events, plotted in Figure \ref{fig:singleplots}. One of these showing very broad \HG{} profile.
\item[GJ 1111 (DX Cnc)] Seven events. Some of these are symmetrical, or combined blue/red wing enhancements.
\item[GJ 1156 (GL Vir)] Seven events. Most of these are enhanced in both the blue and red wing.
\item[GJ 1224] Five events, two stronger plotted in Fig. \ref{fig:singleplots}. In both of the the two stronger events the red wing is enhanced.
\item[GJ 1243] Three events: a slow, a weak, and a broader one. All three showing enhancements in both the blue and red wing.
\item[GJ 1245] Five weak events.
\item[GJ 3622] A few symmetric line profile enhancements.
\item[GJ 3647 (CW UMa)] Two smaller events.
\item[GJ 3971] One symmetric enhancement.
\item[GJ 4053] One symmetric enhancement.
\item[GJ 4071 (V816 Her)] Three symmetric profile enhancements.
\item[GJ 4247 (V374 Peg)] Several flares, three asymmetric events (see \cite{v374peg})
\item[GJ 9520 (OT Ser)] Three weaker events.
\item[HK Aqr] Highly variable line profile, but no line asymmetries.
\end{description}

\label{sect:cmeplots2}
\begin{figure*}
\centering
\includegraphics[width=0.9\textwidth]{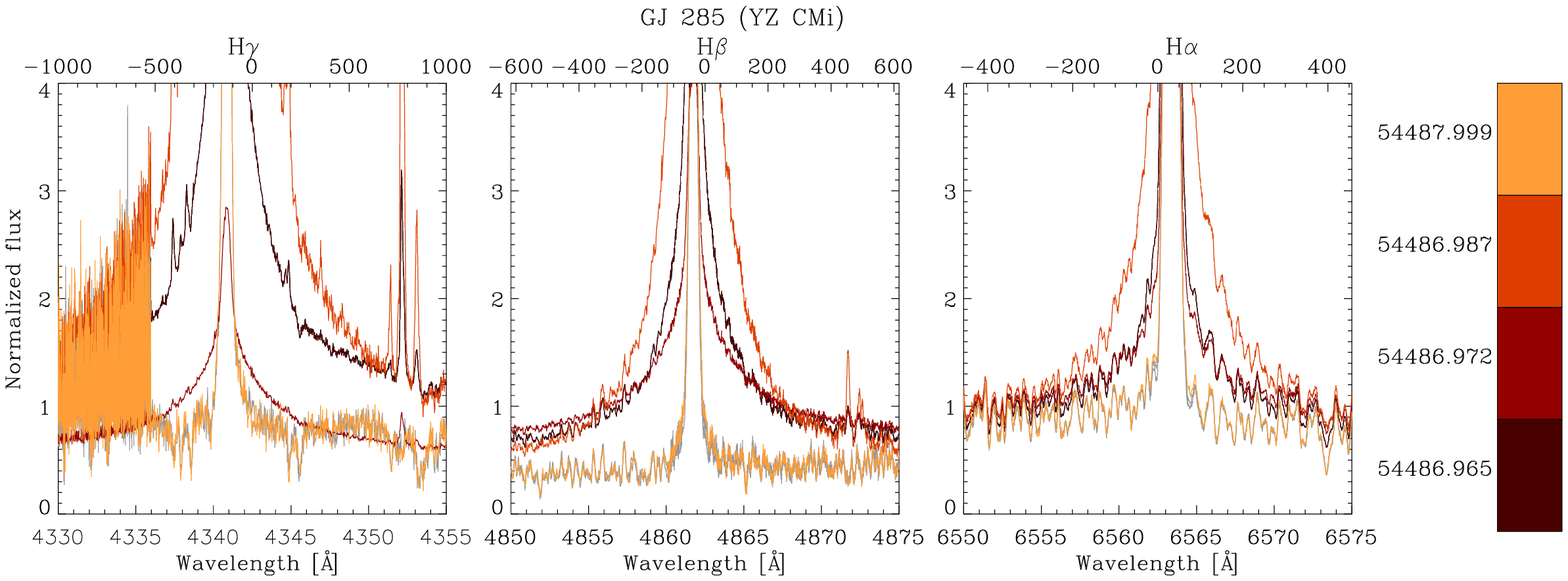}
\includegraphics[width=0.9\textwidth]{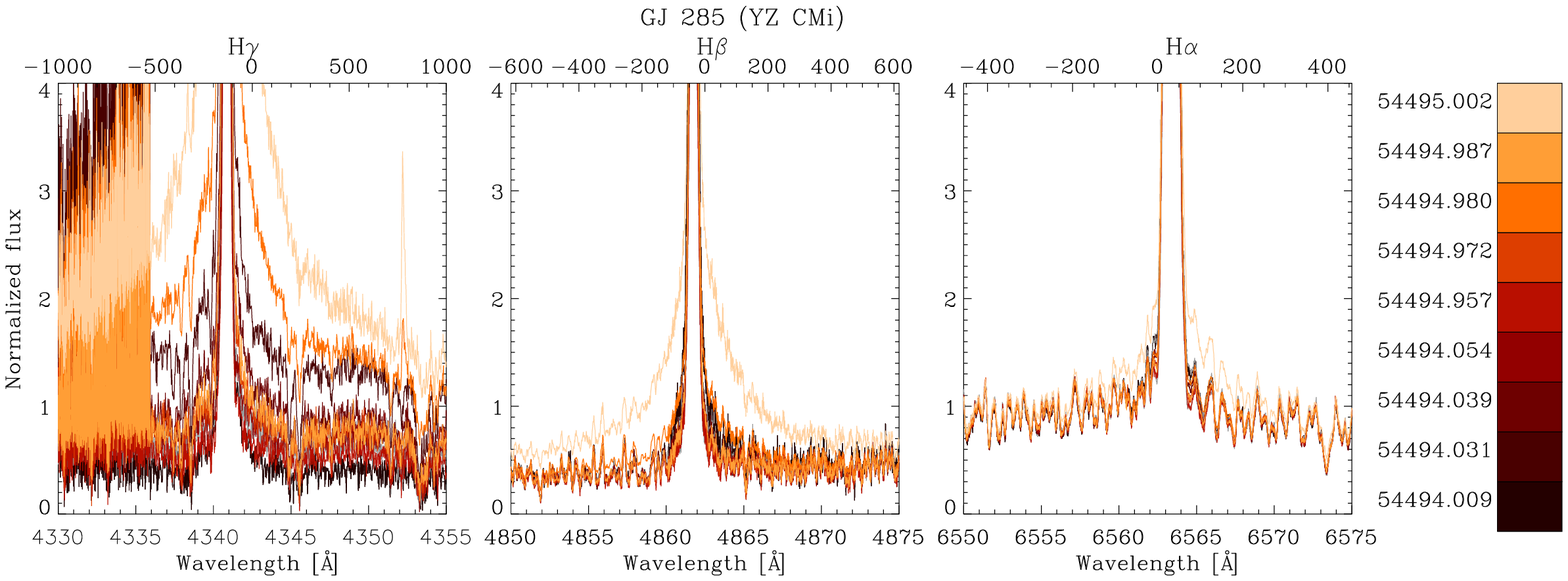}
\includegraphics[width=0.9\textwidth]{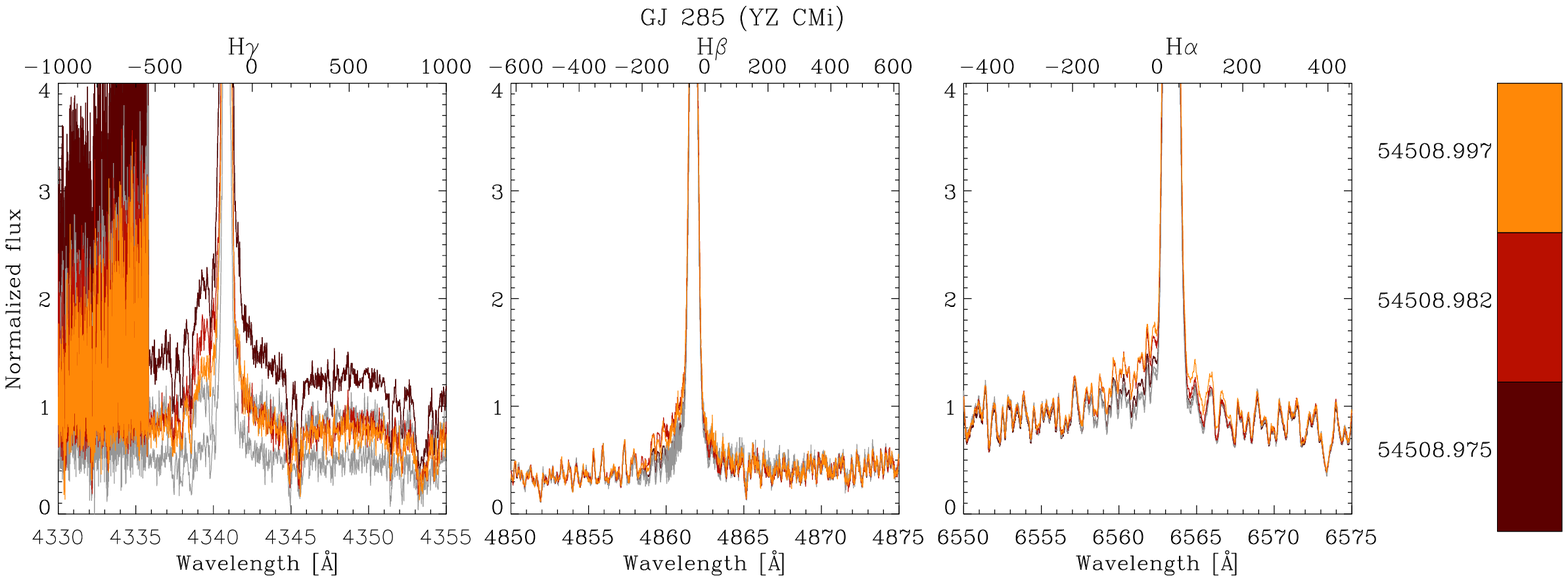}%
\caption{Fig. \ref{fig:cmeplots1}, continued.}
\label{fig:cmeplots2}
\end{figure*}

\begin{figure*}
\centering
\includegraphics[width=0.9\textwidth]{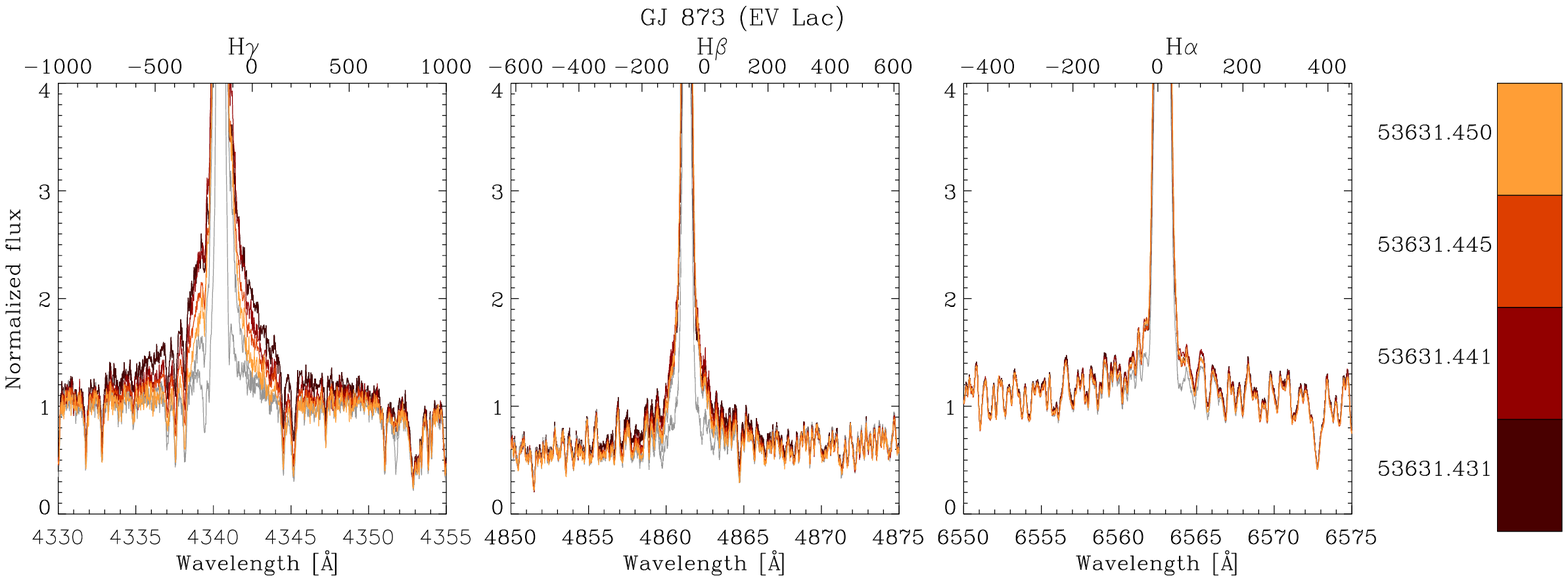}
\includegraphics[width=0.9\textwidth]{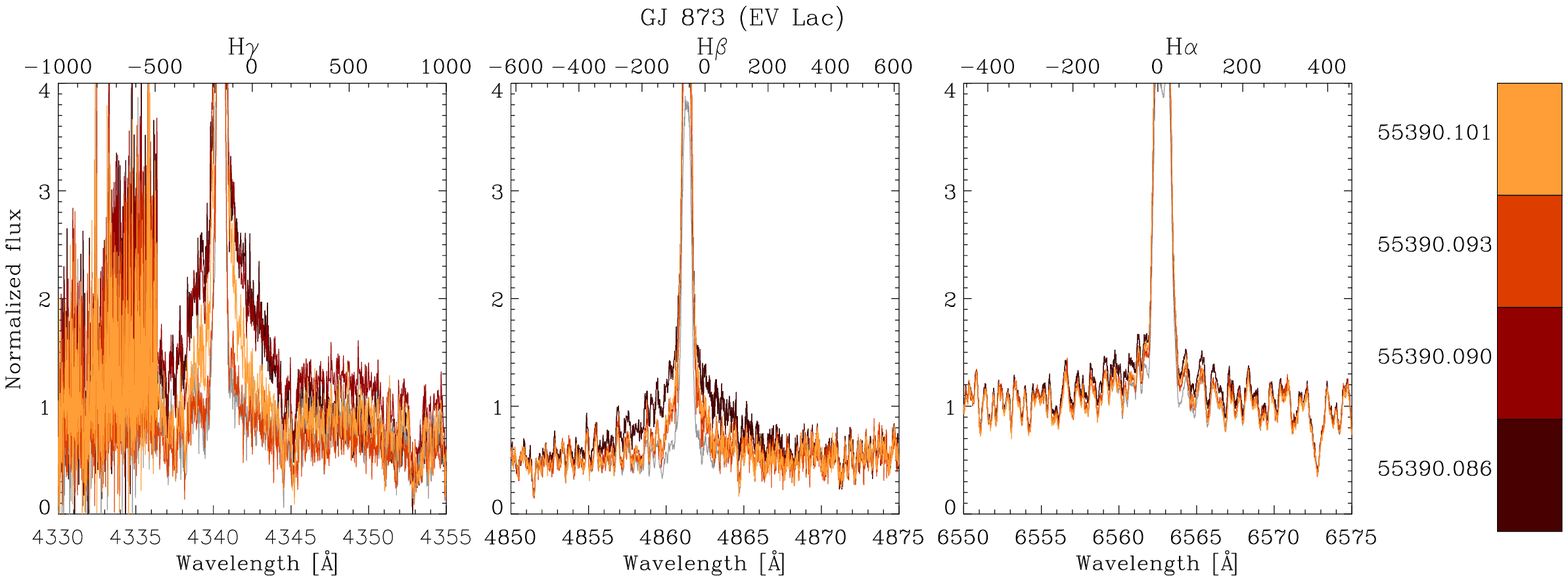}
\includegraphics[width=0.9\textwidth]{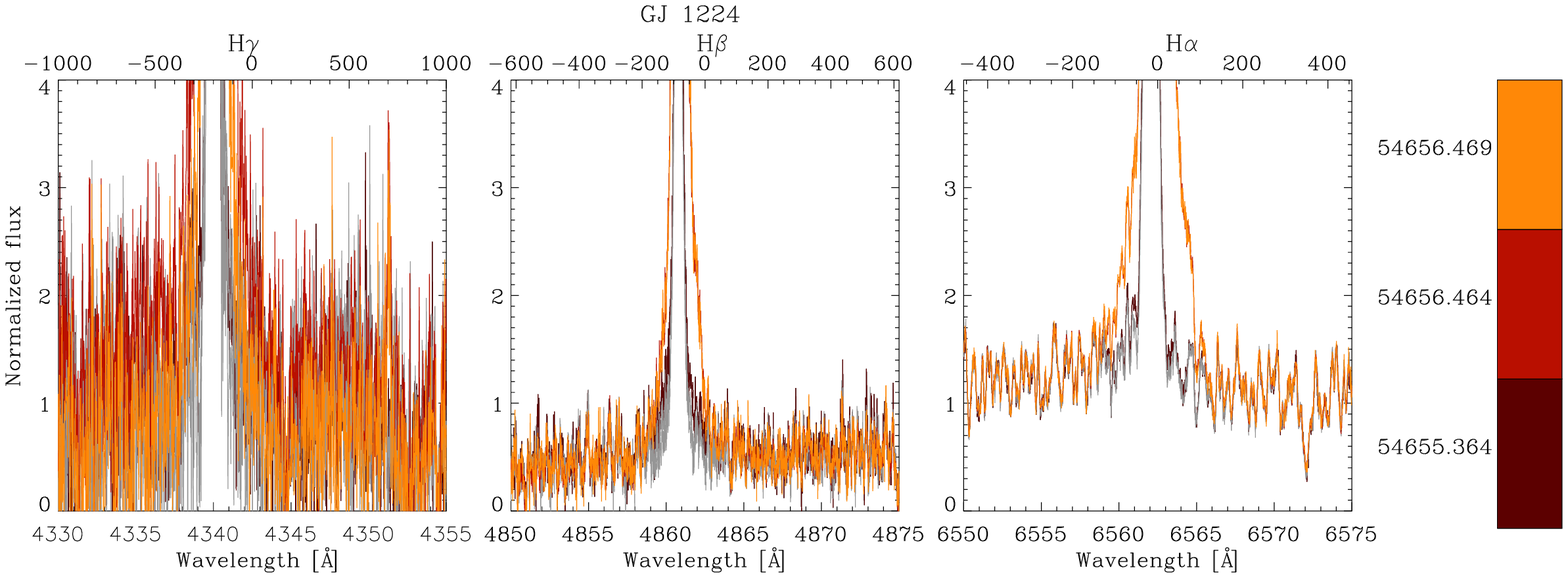}%
\caption{Fig. \ref{fig:cmeplots1} continued.}
\label{fig:cmeplots3}
\end{figure*}

\section{Details on Zeeman--Doppler maps in the literature}
\label{sect:appendix-zdi}
For DT Vir (GJ 494) we only found Zeeman--Doppler imaging (ZDI) maps from 2007 by \cite{zdi-early}, however the CME events occurred in 2012 -- during that time the magnetic configuration could change significantly.  

In case of EV Lac (GJ 873), gaps between the CME detections and the ZDI maps \citep{zdi-mid} were too large: the authors
published maps using data from 2006 August and 2007 July--August, while we detected CMEs in 2005 September, 2008 July and 2010 July. 
The available maps suggest that they both have similar patterns: one active region at the equator, and another at $50^{\circ}$ latitude. However, on a timescale of a year, the magnetic configuration undergoes a significant change.

\cite{zdi-mid} also published ZDI maps of YZ CMi (GJ 285) from 2007 and 2008. 
The latter were recovered from 2007 December--2008 February data, which means, that they coincide with the CMEs that happened between 2008 January--February. 
The authors concluded that the large-scale topology of the magnetic field is quite simple: it is almost axisymmetric and mainly poloidal, and consists of a strong polar active region of negative polarity, while the other hemisphere is covered by the emerging field lines. 

In the case of GJ 51, \cite{zdi-late} published ZDI maps, obtained in 2006, 2007 and 2008 (with no further detail on the epoch of the data), whereas the large CME events were detected 2006 August and 2007 October. The ZDI maps show similar topology at all epochs: it is poloidal and axisymmetric, mainly
composed of a very strong dipole aligned with the rotation axis. In all these three maps the phase coverage was very poor, thus the authors added {\em a priori} information in the process, that strongly prefers axisymmetric solutions. 

The event of V374 Peg (analyzed in detail in \citealt{v374peg}, but no comparison was made with magnetic maps) occurred on 2005 August 20, while the ZDI maps by \cite{zdi-v374peg} were reconstructed using data between 2005 August 19--23. 
The authors also published Doppler maps of the surface. 
They concluded that V374 Peg has a very stable magnetic field (see also photometric data of \citealt{v374peg}), with spottedness of about 2\%. These spots are distributed between latitudes 0--60 degrees, no obvious polar spots are seen. Interestingly both their 2005 August and 2006 August maps suggest that the region between 0.75--1.00 phase of the visible hemisphere has no large active regions (their third map from 2005 September has very poor phase coverage). 
The ZDI maps suggest that the magnetic field of V374 Peg is poloidal and axisymmetric, with the visible hemisphere being mostly covered with positive radial field.

\end{appendix}
\end{document}